\documentclass[conference]{IEEEtran}
\input{talk.mac}

\begin{document}
%
\title{An amendment of the BCS theory of superconductivity}

\author{\IEEEauthorblockN{Drago\c s-Victor Anghel}
\IEEEauthorblockA{Institutul National de C\&D pentru Fizica si Inginerie Nucleara -- Horia Hulubei\\
30 Reactorului Street\\
077125 Magurele, Ilfov, Romania\\
Email: dragos@theory.nipne.ro}
}

\maketitle

\begin{abstract}
Although the BCS theory of superconductivity is a well established theory, we have shown that the phenomenology predicted by this model is much richer than previously believed.
By releasing the constraint 
that the attraction band is symmetric with respect to the chemical potential of the system, we observed that the energy gap may have more than one solution, the quasiparticle imbalance may appear in equilibrium, and the transition between the superconducting and the normal metal phases may be of the first order.
The temperature of the superconductor-normal metal phase transition changes with the asymmetry of the attraction band and if we plot the phase transition temperature vs the chemical potential, we obtain a bell shaped curve, similarly to the superconducting dome, generally formed in high-Tc superconductors, but also in superconductors with narrow conduction bands.

While the pairing interaction is a microscopic characteristic of the system, determined by the effective interactions between constituent quasiparticles, the chemical potential is a macroscopic quantity, which can be changed by external conditions, like doping and pressure.
Furthermore, if the conduction band of the system is narrow, then the attraction band is constrained to the conduction band and the chemical potential is not necessary in the center, as it happens in some of the bands in In-doped Pb$_z$Sn$_{1-z}$ and in MgB$_{2}$.
For these reasons, the constraint that the attraction band is symmetric with respect to the chemical potential may be released.

\end{abstract}

\section{Introduction}

In the BCS theory of superconductivity \cite{PhysRev.108.1175.1957.Bardeen}, electrons of opposite momenta and spins and which have energies close to the Fermi energy, couple to form, below the so called ``critical temperature'' $T_c$, the BCS condensate. The formation of the condensate leads to the phenomenon of superconductivity.
The phase transition that takes place at $T_c$, when the solid goes from the superconducting phase (at temperatures $T<T_c$) to the normal metal phase (at $T>T_c$), is of the second order and is marked--beside the conspicuous jump in resistance--by a jump in the heat capacity. 
Due to the formation of the condensate, the spectrum of the quasiparticle excitations in the superconducting phase shows a temperature dependent energy gap $\Delta(T)$, which decreases monotonically with $T$, from $\Delta(T=0) \equiv \Delta_0$ to $\Delta(T=T_c) = 0$.
The ratio $\Delta_0/T_c$ has an universal value around 1.7, in rather good agreement with experimental data on low temperature superconductors \cite{PhysRev.108.1175.1957.Bardeen, Tinkham:book}.

In isotropic superconductors, the pairing interaction is manifested between electrons which belong to the attraction band, which is a single-particle energy interval, denoted here by $I_V \equiv (\mu-\hbar\omega_c, \mu+\hbar\omega_c)$.
The parameter $\hbar\omega_c$ is of the order of the Debye energy, whereas the center of the attraction band, $\mu$, is associated with the chemical potential of the system (or the Fermi energy).
%
However, in \cite{PhysicaA.464.74.2016.Anghel} it was shown that if the chemical potential of the system is $\mu_R$ and $\mu_R \ne \mu$, then the phenomenology predicted by the model changes dramatically: the energy gap changes and a quasiparticle imbalance appears in equilibrium. Furthermore, not only that the temperature of the superconductor-normal metal phase transition changes, but the phase transition changes qualitatively, becoming of the first order.
If we denote the phase transition temperature by $T_{ph}$, to differentiate it from the BCS critical temperature $T_c$, and plot it vs the difference $\mu_R - \mu$, we observe that it has a maximum at $\mu_R - \mu = 0$ and decreases monotonically with $|\mu_R - \mu|$. For $|\mu_R - \mu| \ge 2\Delta_0$ the superconducting phase does not form anymore, i.e. $T_{ph} = 0$.

The difference $\mu_R - \mu$ is a measure of the asymmetry of the attraction band with respect to the chemical potential. 
The chemical potential may be influenced by pressure or doping, so if we assume that the asymmetry has a monotonic dependence on any of these parameters, then the phase transition temperature may form a maximum when plotted against that parameter, like the 
superconducting dome, generally observed in high-$T_c$ superconductors.
Furthermore, in certain materials, like MgB$_2$ \cite{PhysRevLett.87.047001.2001.Bouquet, PhysRevLett.87.137005.2001.Szabo} and In-doped Pb$_z$Sn$_{1-z}$ \cite{LowTempPhys.41.112.2015.Parfeniev}, the extremum of at least one of the conduction bands lies close enough to the chemical potential to lead to an asymmetric attraction band. In such materials, the asymmetric attraction band changes with pressure or doping.

We are not aware of any single-band superconductors with asymmetric attraction band. Nevertheless, this constitutes our first step in the study of the effects of such an asymmetry on the properties of the superconducting phase and the results, as mentioned above, are significant.
The extension of these results to multi-band superconductors will constitute the subject of another study.

The main result of this paper is the dependence of the phase transition temperature on the asymmetry of the attraction band, $\mu_R-\mu$. Nevertheless, to make the paper more readable we shall introduce in Section~\ref{sec_BCS} the basic concepts and notations, whereas the main result will be presented and discussed in Section~\ref{sec_asym}. The conclusions are presented in Section~\ref{sec_concl}.

%

\section{The standard BCS model} \label{sec_BCS}

Here we introduce briefly the basic notations and concepts.
We denote by $|\bk s \rangle$ the electron's wavefunction, where $\bk$ is the electron's wavevector and $s \equiv \uparrow, \downarrow$ is the electron's spin. If $c_{\bk s}^\dagger$ and $c_{\bk s}$ are the electrons creation and annihilation operators, then the BCS Hamiltonian reads \cite{PhysRev.108.1175.1957.Bardeen,Tinkham:book}
\begin{equation}
  \hat\cH_{BCS} = \sum_{\bk s}\epsilon^{(0)}_\bk c_{\bk s}^\dagger c_{\bk s} + \sum_{\bk\bl} V_{\bk\bl} c^\dagger_{\bk\uparrow} c^\dagger_{-\bk\downarrow} c_{-\bl\downarrow} c_{\bl\uparrow} . \label{def_H_BCS}
\end{equation}
Introducing the notation $b_{\bk} = \langle c_{-\bk \downarrow} c_{\bk \uparrow} \rangle$ (where by $\langle \cdot \rangle$ we denote the average) and replacing $c_{-\bk\downarrow} c_{\bk\uparrow}$ in (\ref{def_H_BCS}) by the equivalent form $b_\bk + (c_{-\bk\downarrow} c_{\bk\uparrow} - b_\bk)$, keeping only the first order terms in the difference $(c_{-\bk\downarrow} c_{\bk\uparrow} - b_\bk)$, one arrives to a Hamiltonian which is quadratic in the operators $c_{\bk s}^\dagger$ and $c_{\bk s}$ and may be diagonalized (see for example \cite{Tinkham:book} for a good introduction).
After diagonalization, one obtains
\begin{equation}
  \hat\cH = \mu \hat N + \sum_\bk(\xi_\bk-\epsilon_\bk+\Delta b_\bk^*) + \sum_\bk\epsilon_\bk(\gamma^\dagger_{\bk 0}\gamma_{\bk 0} + \gamma^\dagger_{\bk 1}\gamma_{\bk 1}) , \label{HM_BCS}
\end{equation}
where $\hat N = \sum_{\bk s} c_{\bk s}^\dagger c_{\bk s} = \sum_{\bk s} n_{\bk s}$ is the particle number operator, $\xi_\bk \equiv \epsilon^{(0)}_\bk - \mu$, $\epsilon_\bk \equiv \sqrt{\xi_\bk^2 + \Delta_\bk^2}$, $\Delta_\bk \equiv \sum_\bl V_{\bk\bl} b_\bl$ is the superconducting energy gap, and $\mu$ is a constant which will be identified with the center of the attraction band.
The quasiparticle creation and annihilation operators introduced in (\ref{HM_BCS}) are
$  c_{\bk\uparrow} = u^*_\bk \gamma_{\bk 0} + v_\bk \gamma^\dagger_{\bk 1} , \ 
  c^\dagger_{\bk\uparrow} = u_\bk \gamma^\dagger_{\bk 0} + v^*_\bk \gamma_{\bk 1} , \ 
  c^\dagger_{-\bk\downarrow} = -v^*_\bk \gamma_{\bk 0} + u_\bk \gamma^\dagger_{\bk 1} , \ {\rm and}\ 
  c_{-\bk\downarrow} = -v_\bk \gamma^\dagger_{\bk 0} + u^*_\bk \gamma_{\bk 1}$,
where $u_\bk$ and $v_\bk$ are
\begin{equation}
  |v_\bk|^2 = 1 - |u_\bk|^2 = \frac{1}{2} \left(1 - \frac{\xi_\bk}{\epsilon_\bk}\right) . \label{def_uv}
\end{equation}
%

To simplify the equations to be able to perform analytical calculations, one in general makes the assumption that $V_{\bk \bl} \equiv -V$ for any $\bk$ and $\bl$, such that both, $\epsilon_\bk^{(0)}, \epsilon_\bl^{(0)} \in I_V = (\mu-\hbar\omega_c, \mu+\hbar\omega_c)$.
In such a case, the energy gap becomes independent of $\bk$,
\begin{eqnarray}
  \Delta &=& - V \sum_\bl \langle c_{-\bk\downarrow} c_{\bk\uparrow}\rangle . \label{def_Delta0}
\end{eqnarray}
Replacing in (\ref{def_Delta0}) the electrons creation and annihilation operators by their expressions in terms of the quasiparticle operators, one arrives to the equation for the energy gap \cite{Tinkham:book}
\begin{equation}
  1 = \frac{V}{2} \sum_\bk \frac{1 - n_{\bk 0} - n_{\bk 1}}{\epsilon_k} , \label{def_Delta2}
\end{equation}
where $n_{\bk i} = \langle \gamma^\dagger_{\bk i}\gamma_{\bk i} \rangle$.

\begin{figure}[t]
  \centering
  \includegraphics[width=7cm,keepaspectratio=true]{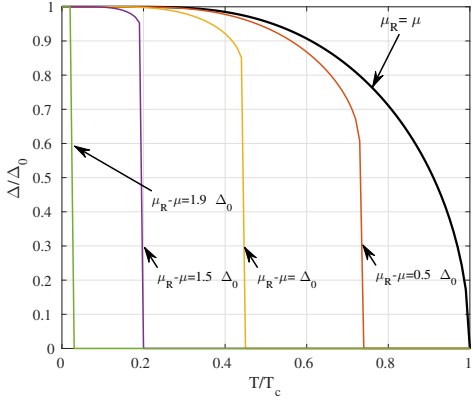}
  \caption{(Color online) The solutions $\Delta$ of Eqs. (\ref{def_F_sigma0_set}) for $(\mu_R-\mu)/\Delta_0 = 0, 0.5, 1, 1.5, 1.9$; $\Delta_0$ is the value of the energy gap at $T=0$, whereas $T_c$ is the BCS critical temperature for a symmetric band.}
  \label{Delta_xF_vs_T}
\end{figure}

Until now, the chemical potential of the system did not enter the formalism.
The standard BCS results are obtained under the assumption that $\mu$ is equal to the chemical potential.
In such a case, the zero temperature solution $\Delta(T=0) \equiv \Delta_0$ is obtained by setting $n_{\bk i} = 0$ for any $\bk$ and $i=0,1$ in (\ref{def_Delta2}). If the electrons single-particle energy spectrum have a constant density of states $\sigma_0$ (for each spin projection), then in the low coupling limit ($\sigma_0 V \ll 1$) $\Delta_0 = 2\hbar\omega_c\exp[-1/(\sigma_0V)]$.
Similarly, by setting $\Delta = 0$ in (\ref{def_Delta2}), one obtains the critical temperature $T_c = (A\hbar\omega_c/k_B) e^{-1/(\sigma_0V)}$, where $A = 2 e^\gamma/\pi \approx 1.13$ and $\gamma \approx 0.577$ is the Euler's constant (see \cite{Tinkham:book} for details).
The solution of (\ref{def_Delta2}) for $T=0$ to $T_c$ is plotted as the upper curve in Fig.~\ref{Delta_xF_vs_T}.


\section{Superconductivity for asymmetric attraction band} \label{sec_asym}

Let us denote the chemical potential of the system by $\mu_R$ and generalize the results from the previous section by assuming that $\mu_R$ may be different from $\mu$. In such a case, the attraction band is not symmetric with respect to the chemical potential $\mu_R$.
To calculate the partition function, we follow \cite{PhysicaA.464.74.2016.Anghel} and write the (average) total particle number as
\begin{eqnarray}
  N &\equiv& \langle\{n_{\bk i}\},\mu|\hat N|\{n_{\bk i}\},\mu\rangle
  = N_0 + \sum_{\bk, i} n_\bk \frac{\xi_\bk}{\epsilon_\bk}
  , \label{N_exp_val1}
\end{eqnarray}
where $N_0 = N' + \sum_{\bk}2v^2_{\bk}$ and $N'$ is the number of particles from outside of the energy interval $[\mu-\hbar\omega_c,\mu+\hbar\omega_c]$. 
Similarly, the total energy of the system is
\begin{equation}
  \cH = E_0 + \sum_\bk\epsilon_\bk (\gamma^\dagger_{\bk 0}\gamma_{\bk 0} + \gamma^\dagger_{\bk 1}\gamma_{\bk 1}) , \label{HM_BCS_eff}
\end{equation}
where
\begin{equation}
  E_0 
  = \mu N + \sum_\bk (\xi_\bk - \epsilon_\bk) + \frac{\Delta^2}{V} . \label{def_E0}
\end{equation}
and $N$ is given by Eq. (\ref{N_exp_val1}).

The partition function is
\begin{eqnarray}
  \ln(\cZ)_{\beta\mu} &=& - \sum_{\bk i} [(1 - n_{\bk i}) \ln(1 - n_{\bk i}) + n_{\bk i} \ln n_{\bk i} ] \nonumber\\
  && - \beta (E-\mu_R N) , \label{cZ_beta_mu}
\end{eqnarray}
where $N$ and $E \equiv \langle \cH \rangle$ are given by (\ref{N_exp_val1}) and (\ref{def_E0}) \cite{PhysicaA.464.74.2016.Anghel}.
Maximizing $\ln(\cZ)_{\beta\mu}$ with respect to the populations $n_{\bk i}$ (see \cite{PhysicaA.464.74.2016.Anghel} for details), we obtain a system of equations which have to be solved self-consistently to determine the energy gap and the populations.
%
%
If the electrons single-particle energy spectrum is $\sigma_0$ (is constant), this system reads
\begin{subequations}\label{def_F_sigma0_set}
\begin{eqnarray}
  \frac{2}{\sigma_0 V} &=& \int_{-\hbar\omega_c}^{\hbar\omega_c} \frac{1 - n_{\xi 0} - n_{\xi 1}}{\epsilon(\xi)} d\xi , \label{Eq_int_Delta1} \\
  n_{\xi i} &=& \frac{1}{e^{\beta[\epsilon(\xi)-(\mu_R-\mu)(\xi - F)/\epsilon(\xi)]}+1} , \ i = 0,1 \label{pop_til_eps_sigma0} \\
  F &\equiv& \frac{ \int_{-\hbar\omega_c}^{\hbar\omega_c} ( 1 - n_{\xi 0} - n_{\xi 1} ) \frac{\xi}{\epsilon^3(\xi)} \, d\xi }
  { \int_{-\hbar\omega_c}^{\hbar\omega_c} \frac{(1 - n_{\xi 0} - n_{\xi 1}) d\xi}{\epsilon^3(\xi)} } . \label{def_F_sigma0}
\end{eqnarray}
\end{subequations}
The set (\ref{def_F_sigma0_set}) is symmetric under the interchange $\mu_R - \mu \to \mu - \mu_R$, $F \to --F$, and $\xi \to -\xi$. 
So, by solving it for $\mu_R - \mu > 0$, we obtain all the solutions, including those for $\mu_R - \mu < 0$.

In Fig.~\ref{Delta_xF_vs_T} we present the solutions for the energy gap, obtained for different values of $\mu_R-\mu$.
We see that if $\mu_R \ne \mu$, the energy gap is smaller than the standard BCS gap--which is the top black curve in Fig.~\ref{Delta_xF_vs_T}--at any temperature and the phase transition temperature $T_{ph}$ is also smaller that the BCS critical temperature $T_c$.
Nevertheless, eventually the most important feature that appears when $\mu_R \ne \mu$ is that the phase transition occurs abruptly, in the sense that the energy gap jumps from a finite value to zero, at $T_{ph}$.

\begin{figure}[t]
  \centering
  \includegraphics[width=7cm,bb=0 0 695 576,keepaspectratio=true]{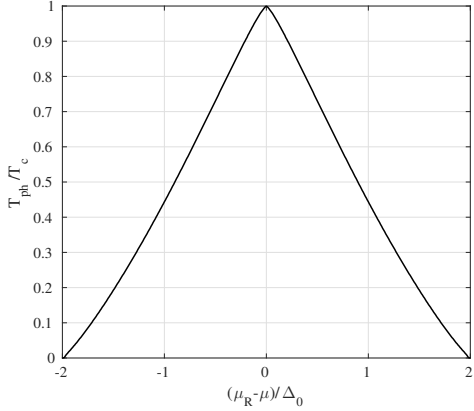}
  \caption{The phase transition temperature vs the asymmetry of the attraction band $\mu_R-\mu$.}
  \label{Tph_vs_yR}
\end{figure}

In Fig.~\ref{Tph_vs_yR} we plot the phase transition temperature vs $\mu_R - \mu$. We see that the function $T_{ph}(\mu_R-\mu)$ has a maximum at $\mu_R = \mu$ and decreases to zero as $|\mu_R-\mu|$ increases to $2\Delta_0$. If $|\mu_R-\mu| \ge 2\Delta_0$, the energy gap cannot be formed anymore and the superconducting phase does not exist.
If the difference $\mu_R - \mu$ varies monotonically with pressure or doping, then $T_{ph}$ plotted vs pressure or doping forms a kind of superconducting dome. A similar behavior was observed also in superconductors with asymmetric attraction band, for example in \cite{LowTempPhys.41.112.2015.Parfeniev}.

If $\mu_R \ne \mu$, then $F \ne 0$ \cite{PhysicaA.464.74.2016.Anghel} and according to (\ref{pop_til_eps_sigma0}) $n_{\xi i} \ne n_{-\xi i}$ and a population imbalance appears \cite{arXiv160907931.Anghel}.
In the standard BCS theory, population imbalance may appear only in nonequilibrium systems \cite{PhysRevLett.28.1363.1972.Clarke, PhysRevLett.28.1366.1972.Tinkham, PhysRevB.6.1747.1972.Tinkham, PhysRevB.22.4346.1980.Smith, PhysRevB.21.3879.1980.Smith}.

\section{Conclusions} \label{sec_concl}

We have briefly reviewed the BCS formalism for asymmetric attraction band proposed in \cite{PhysicaA.464.74.2016.Anghel} and we calculated the temperature of the superconductor-normal metal phase transition for $\mu_R - \mu$ taking values in the interval $[-2\Delta_0, 2\Delta_0]$, where $\Delta_0$ is the energy gap in the standard BCS theory, at zero temperature.
For $|\mu_R - \mu| \ge 2\Delta_0$ the energy gap is zero at any temperature and therefore the superconducting phase cannot exist.
The phase transition temperature is maximum when $\mu_R = \mu$ and decreases monotonically when $|\mu_R - \mu|$ increases.
The phase transition is in general of the order 1, except for the case when $\mu_R = \mu$, when we obtain the standard BCS transition of the second order.
If for a certain material the difference $\mu_R - \mu$ varies monotonically with pressure or doping, then the plot of the phase transition temperature vs these variables lead to a form similar to the superconducting dome or to the bell-shape form obtained in In-doped Pb$_z$Sn$_{1-z}$ \cite{LowTempPhys.41.112.2015.Parfeniev}.

\section{Acknowledgments}

Discussions with Dr. G. A. Nemnes are gratefully acknowledged.
This work has been financially supported by the ANCS (project PN-09370102 PN 16420101/ 2016). Travel support from Romania-JINR Collaboration grants 4436-3-2015/2017, 4342-3-2014/2015, and the Titeica-Markov program is gratefully acknowledged.

\end{document}